\documentstyle[12pt]{article}
\begin{document}
\def\L{{\cal L}_{eff}^{(0)}}
\def\ra{\rightarrow}
\def\beq{\begin{equation}}
\def\eeq{\end{equation}}
\def\bea{\begin{eqnarray}}
\def\eea{\end{eqnarray}}
\def\eg{\hbox{\it e.g.}}
\def\ie{\hbox{\it i.e.}}        \def\etc{\hbox{\it etc.}}
\def\abs#1{\left| #1\right|}
\def\vev#1{\left\langle #1\right\rangle}
\begin{titlepage}
\baselineskip .35cm
\begin{center}
\hfill UCSD/PTH 92/06\\
\vskip 1.5cm
{\large \bf  Light Quark Masses and Quarkonium Decays}
\vskip 1.7cm
{Kiwoon Choi}
\vskip .05cm
{Department of Physics}
\vskip .05cm
{University of California at San Diego}
\vskip .05cm
{9500 Gilman Drive}
\vskip .05cm
{La Jolla, CA 92093-0319, USA}
\vskip 4.5cm
{\bf Abstract}
\end{center}
\begin{quotation}
\baselineskip .56cm
After discussing the intrinsic ambiguity in
determining the light quark mass ratio $m_u/m_d$,
we reexamine the recent proposal that  this ambiguity can be resolved
by applying the QCD multipole expansion
for the heavy quarkonium decays.
 It is observed that,
due to instanton effects, some matrix elements
which have been ignored in previous works  can
give a significant contribution to the decay amplitudes,
which results in  a large uncertainty in the
value of   $m_u/m_d$  deduced from
 quarkonium phenomenology.
This uncertainty
can be resolved only by a QCD calculation of some second order
coefficients in the chiral expansion of the decay amplitudes.

\end{quotation}
\end{titlepage}
\setcounter{page}{2}

\baselineskip 0.77cm
It has been  observed by a number of authors
 [$1-5$] that
second order  corrections   in
 chiral perturbation theory can significantly affect
the estimate of the light quark masses.
In particular,  it was pointed out that
the determination of  $m_u/m_d$  suffers from a large uncertainty
due to the instanton-induced   mass renormalization $\cite{georgi,choi}$:
\beq
M\ra \bar{M}(\omega)\equiv M+{\omega}M_I
\eeq
where
the real matrix $M={\rm diag}(m_u, m_d, m_s)$
denotes the light quark masses in the QCD lagrangian,
$M_I\equiv\frac{1}{4\pi f}({\rm
det}M^{\dagger})(
M^{\dagger})^{-1}=\frac{1}{4\pi f}(m_dm_s, m_um_s, m_um_d)$
is the instanton-induced second order mass
with the pion decay constant $f\simeq 93$ MeV, and
$\omega$ is a dimensionless parameter of order unity.
Most of the previous analyses on $M$ do not distinguish
$M$ from $\bar{M}$, and thus the corresponding results  can be interpreted
as those on  $\bar{M}$ for an arbitrary value of
 $\omega$, which leads to a large uncertainty
 in the extracted value of $m_u/m_d$.

Recently it was argued that  the above mentioned
difficulty can be overcome  by noting that
the instanton-induced mass $M_I$ is distinguished  from
the bare mass $M$ through its
 $\theta$-dependence where $\theta$ denotes the CP
violating QCD vacuum angle $\cite{donoghue}$.
If one keeps the $\theta$-dependence explicitly,
$M_I$ always appears with the phase $e^{i\theta}$
due to the winding number of instantons.
Note that $M$ and $e^{i\theta}M_I$ have the same
transformation under
 the QCD chiral symmetry $SU(3)_L\times SU(3)_R\times U(1)_A$ under which
\beq
M\ra e^{i\alpha}LMR^{\dagger},
\quad \theta\ra
\theta+3\alpha,
\eeq
where  $L\in SU(3)_L$, $R\in SU(3)_R$, and $\alpha$ generates
the anomalous $U(1)_A$ rotation.
Then  one may be able to measure $M$ directly, not $\bar{M}$ involving
an arbitrary unknown parameter $\omega$, by probing
the $\theta$-dependence of the QCD dynamics.
Among quantities that  probe the $\theta$-dependence,
the matrix elements
$\vev{\phi|G\tilde{G}|0}$ ($\phi=\pi^0$  or $\eta$)
were considered in
 ref. $\cite{donoghue}$.
It was then argued  that
 the ratio $R_A\equiv
\frac{\vev{\pi^0|G\tilde{G}|0}}{\vev{\eta|G\tilde{G}
|0}}$
can be reliably determined
 by applying the QCD multipole expansion for the
quarkonium decays $\psi^{\prime}\ra J/\psi+ \phi$,
which would allow us to precisely determine $m_u/m_d$.
In this paper, we wish to reexamine whether the QCD multipole
expansion applied for  the quarkonium decays can provide a way
to determine $m_u/m_d$ without doing any nonperturbative
QCD calculation.
Our result then   confirms the  conclusion
of ref. $\cite{choi1}$, viz. in order to precisely determine $m_u/m_d$,
 one needs to calculate  a QCD matrix element
which receives a potentially large instanton contribution.
Since such a calculation is not available at present,
a rather wide range of $m_u/m_d$, including zero, should be considered
as being  consistent
with our present knowledge of QCD.

To proceed, let us  briefly review  the points that were
discussed   in refs. $\cite{choi,donoghue,choi1}$.
In order to extract information on
$M$, one usually considers    measurable   quantities whose $M$-dependence
can be deduced from
 an effective lagrangian  of hadron fields.
This is not a  severe limitation  since it is hard to imagine
a measurable quantity which may provide  useful information
on $M$ through its $M$-dependence but can not be described
by any  hadronic  effective lagrangian.
To satisfy the Ward identities of the QCD
 chiral symmetry,
the effective lagrangian is required to be  invariant
under the  chiral transformations of the involved hadrons
and   also  those of  $M$ and $\theta$  given in eq. (2).
Since  $M$ and $e^{i\theta}M_I$ have the
same chiral transformation property,
for any  term  in the  effective lagrangian
which is first order in $M$,
\eg \,  ${\cal O}_1=a_i\vev{M\Omega_i}$,
there  exists also   the  second order term
  ${\cal O}_2=b_ie^{i\theta}\vev{M_I\Omega_i}$, where
 $\vev{Z}={\rm tr}(Z)$,
 $a_i$ and $b_i$ are chiral coefficients
which are  calculable within QCD, and
 $\Omega_i$ is a  generic local functional of  hadron fields which is
transformed under the chiral symmetry as $M^{-1}$.
This can be understood within QCD by noting
that $e^{i\theta}M_I$ corresponds to the effective current mass
induced by instantons $\cite{georgi,choi}$.
 For any QCD diagram which contains an insertion
of $M$ and thus gives a contribution to ${\cal O}_1$, one can replace $M$
by the instanton-induced  effective mass
$e^{i\theta}M_I$  to make the new diagram  contribute to  ${\cal O}_2$.
This means that the $M$-dependent part of  the effective lagrangian
can always be written as
\beq
{\cal L}_{eff}(\theta)\supset \sum_i
\left[ a_i\vev{M\Omega_i}+b_ie^{i\theta}
\vev{M_I\Omega_i}+...\right],
\eeq
where the  ellipsis denotes other possible higher order terms.

To be invariant under the anomalous $U(1)_A$ symmetry,
 the effective lagrangian is $\theta$-dependent in general.
One may then  expand  the $\theta$-dependent effective lagrangian
 around $\theta=0$:
\beq
{\cal L}_{eff}(\theta)=\sum_{n=0}^{n=\infty}
 \theta^n {\cal L}_{eff}^{(n)}.
\eeq
Clearly  ${\cal L}_{eff}^{(0)}\equiv{\cal L}_{eff}(\theta=0)$
 describes the normal CP conserving strong interactions
while
 the terms of nonzero $n$  describe the $\theta$-dependence
of the QCD dynamics, including the CP violation due to a
nonzero $\theta$.
In view of the arguments leading to eq. (3),
for any $\L$ that includes the corrections of $O(M^2)$,
one can define a transformation of the form:
\beq
M\ra \bar{M}\equiv {M}+\omega M_I, \quad  b_i\ra b_i-\omega a_i,
\eeq
under which $\L$ is invariant\footnote{
Throughout this paper,
 whenever we say about the IT, it is assumed
that the corrections of $O(M^2)$  are included, while those of
$O(M^3)$ are ignored.}.
The above transformation mixes   the instanton-induced mass $M_I$
 with the bare mass $M$,  and thus
 will be called   as the instanton transformation
(IT) in the following discussion.
Obviously   ${\cal L}_{eff}^{(n)}$ ($n\neq 0$) in eq. (4) are
 {\it not} invariant under the IT.
This is nothing but to
mean that  ${\cal L}_{eff}^{(0)}$ does {\it not}
distinguish
the instanton-induced mass
from the bare one, while ${\cal L}_{eff}^{(n)}$ ($n\neq 0$) distinguish
since they arise from the $\theta$-dependence.
Based on this observation, it was stated in ref. $\cite{choi}$ that
the normal CP conserving strong interactions which are described
by ${\cal L}_{eff}^{(0)}$ always concern
the effective mass $\bar{M}$ with  an unspecified value of $\omega$,
while the CP violating amplitudes or the nonderivative
 axion couplings  probe directly the bare mass $M$.
(Note that up to a small mixing with mesons,
 $\theta$ corresponds to the constant mode of axion in axion
models.)
For instance, if ${\rm det} (M)=0$, then $\theta$ can be rotated
away (even in the case without axion) and  axion becomes massless
 although ${\rm det}(\bar{M})\neq 0$.

  Although ${\cal L}_{eff}^{(n)}$'s ($n\neq 0$) in eq. (4)
might be useful also, one
 usually considers  only   ${\cal L}_{eff}^{(0)}$ in extracting
 information on $M$ from experimental data.
Of course the reason is that
 it is quite nontrivial to find a link between
 the available experimental data
and the terms of nonzero $n$.
The IT was defined as  acting  on $M$ and the second
order chiral coefficients $b_i$'s.
Since  ${\cal L}_{eff}^{(0)}$ is invariant,
all  measurable quantities described by
${\cal L}_{eff}^{(0)}$,
\ie \,  expressed in terms of $M$ and other parameters that appear
in ${\cal L}_{eff}^{(0)}$, are also invariant under the IT.
This can be understood also by observing that
the IT can be considered
as a kind of renormalization group (RG) transformation
associated with the instanton-induced mass renormalization $\cite{choi1}$.
In taking this specific  instanton effects
into account, one can use the following scheme.
For $\theta$ which is fixed to be zero,
the contribution to ${\cal L}_{eff}^{(0)}$
 from   instantons with  size $\rho\leq \mu_I^{-1}$
is  taken into account by redefining
 the chiral expansion parameter
as $M\ra M+\omega M_I$, while
that from   larger instantons
($\rho\geq \mu_I^{-1}$)
appears   in  the second order chiral  coefficients $b_i$.
In this scheme,
 the transformation parameter $\omega$ of the IT,
being naturally of order unity,
can be identified as
$\omega(\mu_I)=\int^{\infty}_{\mu_I}\frac{d\mu}{\mu} \, D(\mu),$
where $D(\mu)$ denotes an appropriately normalized
dimensionless instanton density. Then
 the IT corresponds to a renormalization group
 transformation\footnote{
In considering the IT as a RG transformation, it must be
noted that
the scale $\mu_I$
 is introduced  only for the instanton-induced mass
renormalization,  but not for other kind of instanton
effects.
This isolation of the instanton-induced mass renormalization
from other instanton effects can be easily achieved
 in the dilute  instanton
gas approximation.
Then the conclusion that  measurable quantities deduced from
${\cal L}_{eff}^{(0)}$ are invariant under the IT can be
understood   by
restricting ourselves  to the effects of relatively small
instantons for which the dilute gas approximation
is valid.} changing
$\mu_I$, and
 all  measurable  quantities deduced from ${\cal L}_{eff}^{(0)}$
should be independent of our choice of $\mu_I$, \ie \,
are invariant under the IT.

As was argued previously $\cite{choi1}$,
if we restrict ourselves to $\L$,
  the invariance of $\L$ under the IT
 prevents us from
precisely determining $m_u/m_d$.
All equations  deduced from ${\cal L}_{eff}^{(0)}$
 are {\it covariant} under the IT.
As a result, any quantity which is  sensitive to
the IT, \ie \, its variation under the IT
for $\omega=O(1)$ can be comparable to the expected
central value, can {\it not} be fixed by the invariant
 experimental data.
(Of course the invariant combination of such quantities can be fixed.)
One can {\it not} avoid an uncertainty whose size is characterized
by the variation under the IT.
For  $m_u/m_d$, its variation under the IT is roughly
$\omega m_s/4\pi f$ which can be as large as about 1/2,
implying a rather large uncertainty.

To be more specific, let us consider the case of using the pseudoscalar
meson masses as  experimental input.
The relevant part of the
  effective lagrangian is $\cite{gasser}$
$$
{\cal L}_{eff}^{(0)}\, \supset \, \frac{1}{4}f^2\vev{\chi^{\dagger}U+
\chi U^{\dagger}}+
L_6\vev{\chi^{\dagger}U+\chi U^{\dagger}}^2
$$
\beq
+L_7\vev{\chi^{\dagger}U
-\chi U^{\dagger}}^2+L_8\vev{\chi^{\dagger}U\chi^{\dagger}U+\chi
U^{\dagger}\chi U^{\dagger}},
\eeq
where  $\chi=2B_0M$ for
  $B_0=-\vev{0|\bar{u}u|0}/f^2$  defined at chiral limit,
 and the $SU(3)$-valued
$U$ denotes the  pseudoscalar meson octet.
For the above terms,  the IT of eq. (5)  can be written as
 $\cite{kaplan,leutwyler}$:
\bea
&& m_u\ra m_u+\frac{\omega m_dm_s}{4\pi f}
\quad ({\rm cyclic \, \, in} \, \, u,d,s),
\nonumber \\
&&
L_i\ra L_i-\frac{h_i\omega f}{128\pi B_0},
\eea
where $h_6=h_7=1$ and $h_8=-2$.
Note that $m_s$ can be considered to be
 invariant since its variation
 is negligibly small compared
to $m_s$ for $\omega$ of order unity.
In refs. $\cite{kaplan,gasser}$, it was found  that\footnote{
Throughout this paper,   we will use $m_d/m_s\ll m_s/4\pi f$, and thus
ignore  the corrections suppressed by
either  $m_d/m_s$ or $m_d/4\pi f$,
while  keeping only  those of $O(m_s/4\pi f)$.}
\bea
\frac{m_s^2}{m_d^2-m_u^2}&\simeq&
(M_K^4/M_{\pi}^2)({M_{K^0}^2
-M_{K^+}^2+M_{\pi^+}^2-M_{\pi^0}^2})^{-1},
\nonumber \\
\frac{m_d+m_u}{m_d-m_u}&\simeq& (1+\triangle_M)^2 M_{\pi}^2
(M^2_{K^0}-M^2_{K^+}+M^2_{\pi^+}-M^2_{\pi^0})^{-1},
\eea
where
 $\triangle_M=-32L_7B_0m_s/f^2-c$ for $c\simeq 0.33=O(m_s/4\pi f)$.
These  equations  do not fix $m_u/m_d$, but
give only a curve  on the plane of  ($m_u/m_d,\lambda$)
where $\lambda=m_s/m_d$ for the first equation
and $\lambda=L_7$ for the second.
Note that these   curves are  parametrized by $\omega$,
 and  the uncertainties
in $m_u/m_d$ and $\lambda$ are given by their variations
under the IT for $\omega=O(1)$.
We will encounter the same situation even when other kind of measurable
quantities, \eg \, the baryon masses
and the decay amplitudes for
$\eta\ra 3\pi$, $\psi^{\prime}\ra J/\psi+\pi^0 (\eta)$,
are used in the context of an appropriate form of $\L$.

A definite value of $m_u/m_d$  would be obtained
if one can    choose a specific value of $\lambda$,
 but it is possible only through  a QCD  calculation
of $\lambda$ since $\lambda$ is sensitive to the IT
and thus can not be fixed by experimental
data alone.
Such an attempt was made recently by Leutwyler
$\cite{leutwyler}$
for $\lambda=L_7$.
By invoking $\eta^{\prime}$-dominance,   it was argued that $L_7$ falls
into a rather narrow range of negative values,
 thus ruling out $m_u=0$ in view of the second equation in  (8).
However it has been pointed out later that instantons  significantly
suppress the negative $\eta^{\prime}$-contribution to $L_7$, while enhancing
the positive contribution  from the pseudoscalar octet resonances
$\cite{choi1}$.
This would make $\eta^{\prime}$-dominance for $L_7$  not valid,
and thus results in a large  uncertainty in the value
of $L_7$,  allowing $m_u=0$.
In fact,    any quantity which is sensitive to the IT
 receives a potentially large
contribution from instantons. This  makes the QCD
calculation of $\lambda$ and the precise determination of
$m_u/m_d$ even more nontrivial.

So far, we have argued that  there exists an  intrinsic ambiguity
in determining  $m_u/m_d$  by usual manner.
This ambiguity is due to the invariance of $\L$ under the IT, and
as was pointed out in ref. $\cite{donoghue}$,
  might be resolved by including the
{\it non-invariant}
terms ${\cal L}_{eff}^{(n)}$
($n\neq 0$) in the analysis.
In QCD, one can probe the $\theta$-dependence
through the matrix elements involving the
insertion of the two gluon
 operator $G\tilde{G}$,
 \eg , $\vev{\phi|
G\tilde{G}|0}$ ($\phi=\pi^0$ or  $\eta$)
and $\vev{0|(G\tilde{G})(G\tilde{G})|0}$.
The hadronic realizations
of such matrix elements contain ${\cal L}_{eff}^{(n)}$
($n\neq 0$) in general.
In axion models, $\vev{\phi|G\tilde{G}|0}$ describes the  axion-meson mixing
while  the other matrix element is related to the axion mass.
As a result,
physical processes involving axion may provide information
on these matrix elements, and thus on the bare quark mass $M$.
However it is totally unclear whether the normal
strong interaction data   can also provide any information
on these matrix elements.
 In ref. $\cite{donoghue}$, it was observed that the matrix
element $\vev{\phi|G\tilde{G}|0}$  appears in the QCD
multipole expansion applied for the heavy quarkonium decays
$\Phi^{\prime}\ra\Phi+\phi$ ($\Phi=J/\psi$ or $\Upsilon$).
The major purpose of this paper
is to examine  whether  this allows us to
determine $m_u/m_d$ without doing any nonperturbative QCD calculation.

The quarkonium decays $\Phi^{\prime}\ra\Phi+\phi$
can be described by
the effective lagrangian of eq. (3)
with  $\Omega_i$'s containing the $SU(3)$-singlet heavy quarkonium
fields $\Phi$ and $\Phi^{\prime}$
together with  $\phi$. The resulting decay
amplitudes $H(\phi)$
 (for $\Phi^{\prime}$ at rest) can be written as:
\beq
H(\phi)=\epsilon_{ijk}
\Phi^{\prime}_i\Phi_jp_k\left[\hat{x}\vev{M\phi}+\hat{\gamma}
\vev{M_I\phi}+...\right],
\eeq
where $\Phi^{\prime}_i$ and $\Phi_i$ denote the spin vectors
 of $\Phi^{\prime}$ and $\Phi$ respectively, $p_i$ is the momentum of
 $\phi=\phi^a\lambda^a$.
Clearly  the above amplitudes  must be
invariant under the IT of eq. (5), acting  on
$M$ and also on $\hat{\gamma}$ which is proportional to an appropriate
combination of $b_i$'s.
As was discussed by Voloshin and Zakharov (VZ) $\cite{voloshin}$,
 the QCD multipole expansion whose
expansion parameter is  the inverse of the heavy quark mass
provides another expression for $H(\phi)$:
\beq
H(\phi)=\epsilon_{ijk}\Phi^{\prime}_i\Phi_jX_k(\phi) \, W_0,
\eeq
where $W_0$ depends on the quarkonium wavefunctions, and
\beq
X_i(\phi)\equiv\vev{\phi|g^2E^a_kD_kB_i^a|0}
\equiv p_i X(\phi).
\eeq
At the leading  order in the multipole expansion, the quarkonium
wavefunctions  can be considered to be independent
of $M$, and thus $W_0$ is invariant under the IT.
This means that $X(\phi)$ which is written as
\beq
X(\phi)=x\vev{M\phi}+\gamma\vev{M_I\phi}+...,
\eeq
is invariant under the IT of eq. (5), implying that
$\gamma\equiv\hat{\gamma}/W_0$ is transformed as
\beq
\gamma\ra \gamma-\omega x.
\eeq

In any case, one can define
$$
A_i(\phi)\equiv \vev{\phi|g^2\partial_k(E^a_kB^a_i)|0}=p_i A(\phi),
$$
\beq
B_i(\phi)\equiv - \vev{\phi|g^2B^a_iD_kE_k^a|0}=p_i B(\phi),
\eeq
so that
\beq
X(\phi)=A(\phi)+B(\phi).
\eeq
Note that  $A(\phi)$ can be written as
$A(\phi)=\frac{i}{12}\vev{\phi|g^2G\tilde{G}|0}$ where
$G\tilde{G}\equiv G^a_{\mu\nu}\tilde{G}^{a\mu\nu}$.
In the attempt to determine $m_u/m_d$ using
 the quarkonium decays, the quantities of interests are
 $R_X\equiv X(\pi^0)/X(\eta)$
and $R_A\equiv A(\pi^0)/A(\eta)$.
Then $R_X$ can be fixed by the quarkonium decay data, while it is necessary
to fix $R_A$ to determine $m_u/m_d$.
In fact  $R_A=R_X$ at the leading order in $M$, however
we need to include the corrections of $O(M^2)$ for a
meaningful determination of $m_u/m_d$.
In ref. $\cite{donoghue}$,
following  ref. $\cite{voloshin}$,
it was simply assumed that
 $\abs{B(\phi)}\ll\abs{A(\phi)}$, which would imply  $R_X\simeq R_A$
even at $O(M^2)$.
Here we first argue that
there is no reason for this assumption to be viable,
and later show how the instanton-induced mass renormalization
 promotes $R_A$ to be  greater than $R_X$.

The statement   that $B(\phi)$ may be significantly
smaller than  $A(\phi)$ was first made by VZ
$\cite{voloshin}$ who  observed that $A(\phi)=O(\sqrt{N_c})$, and thus
 is enhanced by one power
of $N_c$ with respect to the naive $N_c$-counting.
This enhancement is due to the $\eta_0$-pole where the $SU(3)$-singlet
 $\eta_0$ denotes  $\eta^{\prime}$ at chiral limit.
Roughly we have
\beq
\vev{\phi|g^2 G\tilde{G}|0}
\simeq \vev{\eta_0|g^2 G\tilde{G}|0}\times
M^2_{\phi\eta_0}\times \frac{1}{M_{\eta_0}^2},
\eeq
where the $\eta_0$-pole ($\equiv 1/M_{\eta_0}^2$) is $O(N_c)$.
Note that
the mass-squared mixing $M^2_{\phi\eta_0}$ between $\eta_0$ and $\phi$
is suppressed by the light quark mass $M$, but is O(1) in the
$N_c$-counting, and
 $\vev{\eta_0|g^2G\tilde{G}|0}=O(1/\sqrt{N_c})$
obeys the naive
$N_c$-counting rule.
In fact, the same  enhancement can occur also for $B(\phi)$  if
$\vev{\eta_0|B_i^aD_kE^a_k|0}$ is nonvanishing.
 Then $B(\phi)$
 can  be equally important as
$A(\phi)$ even in the large $N_c$-limit.
If   $B(\phi)$ does not receive any contribution from the
intermediate $\eta_0$
and thus is $O(1/\sqrt{N_c})$,
we would have
$\abs{{B(\phi)}/{A(\phi)}}\simeq
{M_{\eta^{\prime}}^2}/{\Lambda^2}$,
where $\Lambda$ denotes a typical hadronic scale which is $O(1)$
in the $N_c$-counting.
Clearly then
$X(\phi)$ will be  dominated by $A(\phi)$ in the large $N_c$ limit.
However in the real case of $N_c=3$, this is not necessarily
true since $M_{\eta^{\prime}}$ is large enough to be comparable
to $\Lambda$.  Again $B(\phi)$ can be equally important as
$A(\phi)$.

VZ  noted also that
the equation of motion $D_kE_k^a=igq^{\dagger}\frac{\lambda^a}{2} q$
 gives $B(\phi)$ an additional power of
  the QCD coupling constant $g$.
However it is hard to imagine that
this extra $g$ means  a real suppression of $B(\phi)$ compared to $A(\phi)$.
  First of all, $g$ is essentially of order one, even for the
renormalization point above $m_b$, although the loop
suppression factor
$\alpha_s/4\pi=g^2/16\pi^2\ll 1$.
 Note that the use of the equation of motion has nothing to do with
 the perturbative QCD loop expansion. Furthermore if we consider
the valence quark  contribution (in the sense of the parton
model) to the vacuum to meson
 matrix element  of an $n$-gluon
 operator,  it would
include a factor $g^n$.
Before using the equation of motion, both $A(\phi)$ and $B(\phi)$
involve two-gluon operators.
The equation of motion changes the two-gluon operator
 in $B(\phi)$
 to  $B_i^aq^{\dagger}\lambda^a q$ which includes
only one gluon field.
Then at least for the valence quark contribution,
 $B(\phi)$ is {\it not} higher order in $g$ compared to
$A(\phi)$ since the matrix element of $B_i^aq^{\dagger}\lambda^a q$ will
be enhanced by $g^{-1}$
 compared to the matrix elements of two-gluon operators.

We have argued that there is no a priori reason to expect  that
$B(\phi)$ is significantly smaller than  $A(\phi)$.
One must include $B(\phi)$ in the analysis, and then
 $R_A$ can be significantly different from $R_X$
at $O(M^2)$.
Among the $O(M^2)$ corrections,
 the instanton-induced mass renormalization is of particular
importance for the determination of $m_u/m_d$.
Thus from now on, we will concentrate on how the instanton-induced mass
renormalization affects $A(\phi)$ and $B(\phi)$,
 so that promotes $R_A$ to be greater than $R_X$.
For this purpose, we study
 the IT of the second order chiral coefficients
 $\alpha$ and $\beta$ that
appear in the following   chiral expansion\footnote{
Here we ignore the electromagnetic effects which were shown to
be negligibly small $\cite{donoghue1}$.}:
$$
A(\phi)=a\vev{M\phi}+\alpha\vev{M_I\phi}
+...,
$$
\beq
B(\phi)=b\vev{M\phi}+\beta\vev{M_I\phi}+...,
\eeq
where   the ellipses denote  other possible second order
terms. Note that eqs. (12)  and (15) imply that $A(\phi)$ and $B(\phi)$
 can be expanded as above although these matrix elements can not be
evaluated by using  $\L$ alone.

In order to derive the IT of $\alpha$, let us
express
the physical  meson fields
$\pi^0$ and $\eta$
  in terms of $\phi_3$ and $\phi_8$ as\footnote{In fact,
due to the $SU(3)$ breaking, the $\pi^0$-$\phi_8$  mixing can be different
from the $\eta$-$\phi_3$ mixing at $O(M^2)$. However this difference
does not affect our analysis of the IT, and thus will be ignored.}:
$\pi^0=\phi_3+\epsilon \phi_8,  \quad
\eta=\phi_8-\epsilon\phi_3$.
The mixing parameter  $\epsilon$ can be written as
\beq
\epsilon=\frac{\sqrt{3}}{4}\frac{m_d-m_u}{m_s}
\left[1+\frac{ m_s}
{4\pi f}\kappa\right],
\eeq
where $\kappa$ is a dimensionless parameter of $O(1)$ which represents
the size of the $O(M^2)$-corrections.
 Clearly     $\epsilon$
should be invariant under the IT since
it  describes the physical meson fields
  in terms of the original wavefunctions
$\phi_3$ and $\phi_8$ which are untouched by the IT.
This then gives the following IT of $\kappa$:
\beq
\kappa\ra\kappa+\omega.
\eeq
In fact, using the chiral lagrangian,
 $\epsilon$ was evaluated as
 $\cite{donoghue}$:
\beq
\kappa=-\frac{128\pi B_0}{f}(L_8+3L_7)+ {\rm chiral \, \, logs},
\eeq
assuring the above IT of $\kappa$
(see the IT of $L_i$'s in eq. (7)).

Using eqs. (17) and (18), we can obtain
\beq
A(\pi^0)=\frac{3}{2}(m_u-m_d) a\left[1+\frac{m_s}{4\pi
f}(\frac{\kappa}{3}
-\frac{2\alpha}{3a}+...)\right],
\eeq
while the anomalous Ward identity
$$
\partial_{\mu}(\bar{q}\gamma^{\mu}\gamma_5 M^{-1} q)
=\frac{g^2}{16\pi^2}{\rm tr}(M^{-1})G^{a\mu\nu}\tilde{G}^a_{\mu\nu}
+2\bar{q}i\gamma_5 q
$$
gives
\beq
A(\pi^0)=\frac{4i\pi^2}{3}\frac{m_u-m_d}{m_u+m_d}f_{\pi}
M_{\pi}^2\left[1+O(\frac{m_d}{4\pi f})\right],
\eeq
where $f_{\pi}$ and $M_{\pi}$ denote the decay constant and the  mass of
the physical $\pi^0$ respectively.
Here and in what follows,
 ellipsis  always denotes  a dimensionless
coefficient of $O(1)$ which is untouched by the IT.
Then comparing  (22) with (21) together with
  the expressions of $f_{\pi}$ and $M_{\pi}$
given in ref. $\cite{gasser}$, we find
\bea
a&=&\frac{8i\pi^2B_0f}{9},
\nonumber \\
\frac{\alpha}{a}&=&-\frac{64\pi B_0}{f}(L_8+3L_7+3L_6)+...,
\eea
which yield the following IT
\beq
\alpha\ra \alpha +2\omega a, \quad \quad \beta\ra \beta-\omega (b+3a).
\eeq
Here the IT of $\beta$ is derived from that of $\alpha$ and $\gamma$,
using
$x=a+b$ and $\gamma=\alpha+\beta$.

We are now ready to see how the instanton-induced mass renormalization
affects $R_A$ and $R_X$, and  what is still necessary
to determine $m_u/m_d$  using the QCD multipole expansion
applied for the quarknonium decays.
Using eqs. (12), (17), and (18), we  can obtain
\bea
R_X&\equiv&\frac{X(\pi^0)}{X(\eta)}=\frac{3\sqrt{3}}{4}\frac{m_d-m_u}
{m_s}\left[1+\frac{m_s}{4\pi f} \xi_X\right]
\nonumber \\
R_A&\equiv&\frac{A(\pi^0)}{A(\eta)}=\frac{3\sqrt{3}}{4}
\frac{m_d-m_u}{m_s}\left[1+\frac{m_s}{4\pi f}\xi_A\right],
\eea
where
$\xi_X={\kappa}/{3}-{2\gamma}/{3x}+..., \quad
\xi_A={\kappa}/{3}-{2\alpha}/{3a}+...$
are dimensionless coefficients of $O(1)$.
Then the IT of $\alpha$, $\gamma$, and $\kappa$ derived so far
 gives the following IT:
\beq
\xi_X\ra \xi_X+\omega, \quad
\xi_A\ra \xi_A-\omega.
\eeq
We have argued that the IT can be interpreted as a kind of
RG transformation changing the scale $\mu_I$ that appears
associated with the instanton-induced mass renormalization.
With this interpretation,
the IT of $\xi_{A,X}$ for a maximal value of  the transformation
parameter, \ie \,
$\omega=\omega_{\rm max}$,
represents the negative of the full instanton contribution to
 $\xi_{A,X}$.
Then $\xi_A$ receives  a positive contribution $\omega_{\rm max}$
 from instantons, while
$\xi_X$ receives a negative one with the same size.
Although its precise value is quite sensitive to the unknown
nonperturbative QCD dynamics,
it has been observed that $\omega_{\rm max}$ can be as large as
$2\sim 3$ in the semiclassical instanton
gas approximation $\cite{georgi,choi}$.
Then although not very reliable, for
 such a large $\omega_{\rm max}$,  it is more likely that
$R_A$ is significantly {\it greater} than $R_X$.

Applying the QCD multipole expansion
for   the measured    quarkonium
decay widths ${\Gamma(\psi^{\prime}\ra J\psi + \phi)}$ ($\phi=\pi^0$
or $\eta$), one finds
  $R_X\simeq 4.3\times 10^{-2}$
$\cite{donoghue,ioffe,ioffe1}$.
Also in ref. $\cite{donoghue}$, the size of
  $\hat{R}_A\equiv (m_d+m_u)R_A/(m_d-m_u)$ was  estimated
within the  second order  chiral perturbation theory of the  light
pseudoscalar  mesons, which yields $\hat{R}_A\simeq 8\times 10^{-2}$.
Note that both $R_X$ and $\hat{R_A}$ are
invariant under the IT, and thus can be
fixed  by experimental data.
If we assume  as in ref.
$\cite{donoghue}$ that $\abs{B(\phi)}\ll\abs{A(\phi)}$ which
implies $R_A\simeq R_X$ or equivalently $\xi_A\simeq \xi_X$, the
measured values of
 $R_X$ and $\hat{R}_A$ would give  $m_u/m_d\simeq 0.3$.
However we already argued   that  there is no reason
for $\abs{B(\phi)}\ll\abs{A(\phi)}$. Furthermore
once we include  $B(\phi)$ in the analysis as it must be,
 instantons give a  potentially large
{\it positive} contribution $\omega_{\rm max}$
 to $\xi_A$, while $\xi_X$ receives
a {\it negative} contribution with the same size.
This is essentially due to the instanton-induced $O(M^2)$ piece
in $B(\pi^0)$, \ie \, the term with the coefficient $\beta$ arising from
 the instanton-induced mass renormalization.
Then what we  obtain from the entire analysis can be summarized by
the equation:
\beq
\frac{m_d-m_u}{m_d+m_u}\simeq 0.54 \, (1+\frac{m_s}{4\pi
f}(\xi_A-\xi_X)),
\eeq
where $\delta\xi\equiv(\xi_A-\xi_X)$ is a totally unknown
coefficient of $O(1)$.
If  $\omega_{{\rm max}}=2\sim 3$ whose possibility  was assured
within the semiclassical instanton gas approximation $\cite{georgi,choi}$,
it is  conceivable to assume that
 $\delta\xi$
is dominated by the {\it positive}
instanton contribution $2\omega_{{\rm max}}$, implying
$\xi_A>\xi_X$.
Then   the measured values of $R_X$ and $\hat{R}_A$
would give $m_u/m_d< 0.3$.
In any case,
in order for  $m_u/m_d$ to be precisely determined,
 we still need to compute (within QCD)
 the chiral coefficient
$\delta\xi$ which is very sensitive to the IT

To conclude, we have examined whether
 the QCD multipole expansion applied for the quarkonium decays
$\Phi^{\prime}\ra\Phi \, \phi$ can be useful for the precise
determination of $m_u/m_d$.
The matrix element $B(\phi)$ which has been ignored in previous works
 can significantly
affect the estimate of $m_u/m_d$,
particularly through the $O(M^2)$-piece  induced by instantons.
As a result, a rather wide range of $m_u/m_d$ (including zero)
 can be consistent with
the observed quarkonium decays.
As was concluded in ref. $\cite{choi1}$,
 to determine $m_u/m_d$
precisely,
we  need a QCD calculation of  the  chiral coefficient
$\delta\xi=(\xi_A-\xi_X)$ which is  sensitive to the IT.
If  instanton contribution to $\delta\xi$
dominates over other contribution, which is  conceivable
 for  $\omega_{\rm max}=2\sim 3$,
and thus $\delta\xi>0$,
the observed quarkonium decays imply $m_u/m_d<0.3$, allowing the massless
up quark scenario  $[1,2,3]$
 for the absence of CP violation in strong
interactions.

\begin{center}
{\bf Acknowledgements}
\end{center}
I thank  A. Manohar for bringing my attention
to ref. $\cite{donoghue}$ and also for discussions.
This work was supported  by the DOE contract \#DE-FG03-90ER40546.

\pagebreak

\end{document}